\documentclass[superscriptaddress,twocolumn,amsmath,amssymb,prl, aps]{revtex4}

\usepackage{graphicx}
\usepackage{array}
\usepackage{amsfonts}
\usepackage{amsmath}
\usepackage{amssymb}
\usepackage{fancyhdr}

\newcommand{\Pbo}{$\mathrm{P}_\mathrm{b0}$}

\newcommand{\sio}{$\mathrm{SiO}_\mathrm{2}$}
\newcommand{\Phs}{$\mathrm{^{31}P}$}
\newcommand{\Tplus}{$|T_+\rangle$}
\newcommand{\Tminus}{$|T_-\rangle$}

\begin{document}
\title{Electrical detection of coherent \Phs~spin quantum states}
\author{Andre R. Stegner}
\affiliation{Walter Schottky Institut, Technische Universit\"at
M\"unchen, Am Coulombwall 3, 85748 Garching, Germany}
\author{Christoph Boehme}
\affiliation{University of Utah, Physics Department, 115S 1400E,
Salt Lake City, Utah 84112, USA}
\author{Hans Huebl}
\affiliation{Walter Schottky Institut, Technische Universit\"at
M\"unchen, Am Coulombwall 3, 85748 Garching, Germany}
\author{Martin Stutzmann}
\affiliation{Walter Schottky Institut, Technische Universit\"at
M\"unchen, Am Coulombwall 3, 85748 Garching, Germany}
\author{Klaus Lips}
\affiliation{Hahn-Meitner-Institut Berlin, Kekul\'estr.~5,
12489 Berlin, Germany}
\author{Martin S. Brandt}
\affiliation{Walter Schottky Institut, Technische Universit\"at
M\"unchen, Am Coulombwall 3, 85748 Garching, Germany}

\date{\today}

\begin{abstract}
In recent years, a variety of solid-state qubits has been
realized, including quantum dots \cite{Elzerman:2004, Petta:2005}, superconducting tunnel junctions \cite{Mooij:1999,
Pashkin:2003} and point defects \cite{Kennedy:2002, Jelezko:2004}.
Due to its potential compatibility with existing microelectronics, the
proposal by Kane \cite{Kane:1998, Kane:2000} based on phosphorus
donors in Si has also been pursued intensively \cite{Vrijen:2000,
Hollenberg:2004, Clark:2003}. A key issue of this concept is the
readout of the \Phs~quantum state. While electrical measurements of
magnetic resonance have been performed on single
spins~\cite{Xiao:2004, Brandt:2004}, the statistical nature of these experiments based on random telegraph noise measurements has impeded the readout of single spin states. In this letter, we
demonstrate the measurement of the spin state of \Phs~donor electrons in silicon and the observation of Rabi flops by purely electric means, accomplished by coherent manipulation of spin-dependent charge carrier recombination between the \Phs~donor and paramagnetic localized states at the Si/\sio~interface via pulsed electrically detected magnetic resonance. The electron spin information is shown to be coupled through the hyperfine interaction with the \Phs~nucleus, which demonstrates the feasibility of a recombination-based readout of nuclear spins.
\end{abstract}

\maketitle Since the detection of single charges has become technically
straight forward, it is widely believed~\cite{Kane:1998,
Kane:2000,Hollenberg:2004, Clark:2003,Xiao:2004} that the
realization of spin-to-charge transfer is the key prerequisite for
a successful implementation of single spin phosphorus ($\mathrm{^{31}P}$)
readout devices, capable of determining the actual spin state (spin up $|\negmedspace \uparrow \rangle$ or spin down $|\negmedspace \downarrow \rangle$). Different approaches to the electrical spin readout of \Phs-donor electron
spins have been proposed based on spin-dependent transitions
between neighboring \Phs~atoms~\cite{Kane:1998,
Kane:2000,Hollenberg:2004, Clark:2003}. Since the states involved are
energetically degenerate, these spin-to-charge transfer
schemes are rather difficult to realize. Alternatively,
spin-dependent transitions involving dissimilar paramagnetic
states might be easier to detect as proposed by Boehme and
Lips~\cite{Boehme:2002}. Spin-dependent charge carrier transport
and recombination are known at least since
1966~\cite{Schmidt:1966}, when Schmidt and Solomon already observed
spin-dependent recombination involving \Phs~donors in silicon.
However, it was not demonstrated until 2003~\cite{Boehme:20032}
that the much more sensitive electrical detection of spins via resonant changes of recombination processes is also able to reflect coherent spin motion, which is necessary for a readout of the spin quantum state as opposed to a mere detection of the presence of spins.

\begin{figure}
\centering
\includegraphics[width=0.35\textwidth]{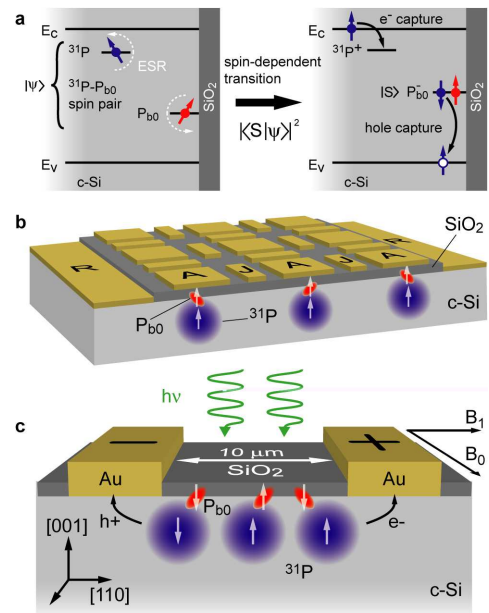}
\caption{(a) Sketch of the readout of \Phs~spin quantum states via spin-dependent charge carrier recombination at \Pbo~centers at the crystalline Si/\sio~interface. The electrons at the \Phs~donor and the \Pbo~center form a weakly-coupled spin pair $|\Psi\rangle$. Since the negatively charged $\mathrm{P}_{\mathrm{b}0}^-$-state is diamagnetic the recombination probability is proportional to the singlet content $|\langle S|\Psi\rangle|^2$ of the initial pair. Electron spin resonance of either constituent of the pair can therefore influence the recombination of excess charge carriers. (b) Incorporation of the \Phs-\Pbo~readout scheme into an array of \Phs~qubits coupled by J gates. Readout of a specific \Phs~qubit is achieved by selective application of a positive gate bias to the corresponding A gate. (c) Sample structure used in the experiments to demonstrate electrical detection of Rabi oscillations of \Phs~qubits. To preferentially locate \Phs~in the vicinity of c-Si/\sio-states, the active sample is a 15~nm thick layer of c-Si, doped with $10^{17}~\mathrm{cm}^{-3}$ \Phs~atoms, on a 5~$\mu$m intrinsic buffer epitaxially grown by low-pressure chemical vapor deposition on a 30~$\Omega \mathrm{cm}$ Boron-doped Si(001) wafer. A high density of \Pbo~centers is achieved by a native oxide. The contacts are a grid structure consisting of 10~nm thick chromium and 100~nm thick gold films and comprising 112 digits of 2~mm length, $10~\mu \text{m}$ width and $10~\mu\text{m}$ distance.
}
\label{fig:schematic}
\end{figure}

Figure~\ref{fig:schematic}(a) illustrates the readout scheme based on spin-dependent recombination demonstrated in this paper. In order to probe \Phs~donor electron spins, spin-dependent excess charge carrier recombination through so-called
$\mathrm{P}_\mathrm{b0}$ centers is used. \Pbo~centers are trivalent Si atoms at the
interface between crystalline silicon (c-Si) and silicon dioxide (\sio) that introduce localized, paramagnetic states
in the Si band gap and dominate electron trapping and
recombination at the interface \cite{Poindexter:1981,
Stesmans:1998}. If a neutral \Pbo~center is located in the vicinity
of a \Phs~donor, the electron bound to the \Phs~and the
electron in the \Pbo~center can form a coupled spin pair
\cite{KSM:1978,Boehme:2003}, whose wave function $|\Psi\rangle$
can exist in an arbitrary superposition of any of its four energy
eigenstates. Two of these states,
$|T_+\rangle = |\negmedspace \uparrow \uparrow \rangle$ and $|T_-\rangle = |\negmedspace \downarrow \downarrow \rangle$, where the first arrow indicates the state of the \Phs~spin and the second arrow the state of the \Pbo~spin, have
triplet symmetry and are independent of the strength of the spin-spin coupling. The remaining two states depend on the coupling: For strong spin-spin interaction, they are the pure triplet $|T_0\rangle = \tfrac{1}{\sqrt{2}}\left(|\negmedspace \uparrow \downarrow \rangle + |\negmedspace \downarrow \uparrow \rangle \right)$ and the singlet
$|S\rangle = \tfrac{1}{\sqrt{2}}\left(|\negmedspace \uparrow \downarrow \rangle - |\negmedspace \downarrow \uparrow \rangle \right)$ states, whereas in the absence of spin-spin coupling, they are product states $|\negmedspace \uparrow \downarrow \rangle$ and $|\negmedspace \downarrow \uparrow \rangle$ which can be represented by equal mixtures of $|T_0\rangle$ and $|S\rangle$. For the intermediate coupling regime, the energy eigenbase changes continuously between the two limiting cases. Once a spin pair is generated, it can relax by a transfer of the electron from the \Phs~donor to the \Pbo, forming a negatively charged $\mathrm{P}_{\mathrm{b}0}^-$. This relaxation transition depends strongly
on the initial symmetry of the two spins within the pair.
Since the energetically lowest doubly occupied charged state of the \Pbo~has to be
diamagnetic due to the Pauli exclusion principle, the relaxation
probability is proportional to the singlet content $|\langle
S|\Psi\rangle|^2$ of the pair~\cite{Boehme:2003}. After subsequent
capture of an excess hole by the $\mathrm{P}_{\mathrm{b}0}^-$, and an excess
electron by the $\mathrm{^{31}P^+}$, both centers return to their
initial uncharged states. This capture of mobile charge carriers e.g. generated by illumination renders the spin-dependent process detectable by current measurements.

The incorporation of this readout scheme into an array of \Phs~qubits is sketched in Fig.~\ref{fig:schematic}(b). While coupling between neighboring qubits is achieved via J gates controlling exchange, A gates are used to couple the \Phs~qubits to the \Pbo~readout centers. Selective readout of a single \Phs~ in this array is achieved by applying a negative gate potential to all A gates except one which is decoupling \Phs~and \Pbo. At one qubit, a positive bias is applied, so that spin-dependent recombination exclusively takes place there, monitored by a current flowing between the two R contacts, which depends on the spin state of the qubit addressed. For an application of spin-dependent recombination to spin readout, the \Phs~-\Pbo~mechanism discussed above might be superior to Kane's original approach~\cite{Kane:1998}. Since \Pbo~states have a much higher localization (more than 80\% of the wave function can be found within a c-Si lattice
constant~\cite{brow:1983}) than the \Phs~donors~\cite{Koiller:2006}, one can expect that gate potentials might influence the \Pbo~wave function to a much lesser degree than the donor wave function, conceptually simplifying the gate-controlled exchange interaction. Note that a sufficiently slow application of a positive potential on an A gate would not only increase the \Phs-\Pbo~exchange, it would also adiabatically reduce the \Phs~hyperfine interaction~\cite{Kane:1998, Tyryshkin:2006}. Since the nuclear spin
affects the \Phs-donor electron spin resonance due to the
hyperfine coupling, the slow application of the
positive bias would adiabatically encode the \Phs~nuclear spin
state into the permutation symmetry of the \Phs-\Pbo~electron
spin pair, while, at the same time, the \Phs~nucleus would become
decoupled. Thus, also a selective and relaxation free \Phs~nuclear spin-readout determined by spin-dependent \Phs-\Pbo~transitions could
be possible via selective application of an electric
field at an individual \Phs~site in the array in Fig.~\ref{fig:schematic}(b).

To demonstrate the feasibility of this readout scheme mapping the \Phs~donor electron spin state on electrical current, pulsed electrically detected magnetic resonance experiments (pEDMR) were performed on the test structure shown in Fig.~\ref{fig:schematic}(c). A resonant microwave pulse is used to induce coherent manipulation of the electron spins by electron spin resonance (ESR). The resulting change of the recombination rate is then detected by a transient photoconductivity measurement after the coherent excitation is turned off~\cite{Boehme:2003}. For the preparation of the initial state of the ensemble of spin pairs, the sample is allowed to attain a low temperature $T=5~\mathrm{K}$ steady state in the presence of a
strong offset photocurrent $I = 50~\mathrm{\mu A}$, a constant magnetic field $B_0 \approx 350~\mathrm{mT}$ and illumination
with a tungsten light source. In this way, a high density of \Tplus~and \Tminus~spin pairs is generated, as the steady state
recombination leads to a depletion of the short lived pair states $|S\rangle$ and, in the case of intermediate or weak coupling, $|T_0\rangle$. pEDMR is induced by an intense, coherent ESR microwave pulse generated by a Bruker E580 pulse bridge and amplified by a 1 kW travelling wave tube amplifier, which imposes a unitary transformation on the spin pair eigenstates. The final state of this transformation is a coherent non-eigenstate, determined by adjustable pulse parameters such as intensity, pulse length and microwave frequency. If this state has an increased singlet content, the recombination rate after the pulse is also increased and will only gradually return to its steady state, as recombination processes of the charge carrier pairs typically take place on a much slower timescale than their manipulation by the strong microwave pulse.
\begin{figure}
\centering
\includegraphics[width=0.35\textwidth]{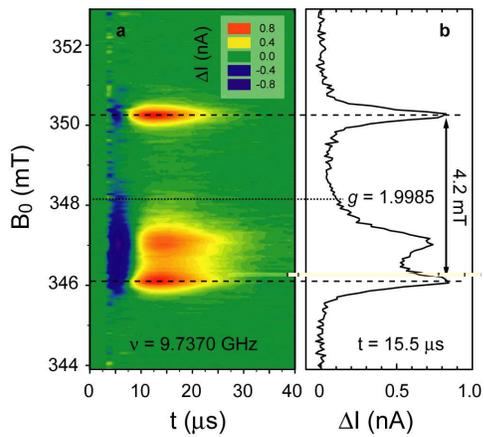}
\caption{Contour plot (a) of the transient current change after a 480~ns long microwave pulse
with a power $P_{\mathrm{MW}} = 1$ W as a function of the
magnetic field $B_0$ and after subtraction of microwave artifacts. To protect the amplifier against overload, the detection system is activated only after $3~\mu \mathrm{s}$. Every time slice has the shape of an ESR spectrum as exemplarily
shown for $t = 15.5~\mu \mathrm{s}$ (b). The spectrum is a superposition of the hyperfine-split resonance of the \Phs~donors and the resonance of the \Pbo~centers.} \label{fig:transients}
\end{figure}
Figure~\ref{fig:transients}(a) displays a contour plot of the change
$\Delta I$ of the photocurrent from its steady state value as a
function of time $t$ after the microwave pulse was turned off. The data shown
were obtained for different magnetic fields $B_0$ and represent the measured data
after subtraction of non-resonant microwave artifacts. The
contour plot displays three resonances at $B_0 =
346.2~\mathrm{mT}$, $B_0 = 347.1~\mathrm{mT}$, and $B_0 =
350.3~\mathrm{mT}$, also visible in Fig.~\ref{fig:transients}(b) which shows a
time slice through the contour plot for $t = 15.5~\mu \mathrm{s}$.
The two features indicated by dashed lines have equal amplitudes,
are separated by $4.2~\mathrm{mT}$ and correspond to a central
$g$-value of 1.9985. This is the characteristic ESR hyperfine signature of
the \Phs~donor electron in silicon \cite{Feher:1959, Young:1997}. The broader
peak at $B_0 = 347.1~\mathrm{mT}$ is an unresolved superposition of signal contributions from $\mathrm{P}_\mathrm{b0}$ centers that have different orientations with respect to the Si (100) surface, which results in resonances at $g = 2.0039$ and $g = 2.0081$ for the orientation of the sample with respect to the magnetic field orientation indicated in Fig.~\ref{fig:schematic}(c) \cite{Poindexter:1981, Stesmans:1998}. The current transients at the three resonances exhibit identical dynamics: An initial strong photocurrent decrease caused by the higher singlet content followed by a slower increase of the current originating from the quenching of the number of triplet states which also have a finite recombination probability~\cite{Boehme:2003}. We conclude from the identical time dependence that the electronic transitions which cause the observed signals must belong to the same recombination process, involving \Phs~donors and \Pbo~centers, as illustrated in Fig.~\ref{fig:schematic}.

The transient response of the photocurrent after the microwave pulse is proportional to the relative singlet density change induced by the microwave pulse~\cite{Boehme:2003}. Therefore, a current change, integrated over a certain time in a box-car type of experiment for the purpose of noise reduction yielding a charge $Q$, reflects the spin state densities directly after the pulse. Since a change of the pulse length $\tau$ causes a different nutation angle for the \Phs~spin on resonance and thus a different singlet content for the \Phs-\Pbo~spin pair after the pulse, a measurement of $Q(\tau)$ must reveal the coherent Rabi nutation of the spins in resonance.

To observe Rabi flops of the \Phs~spins, the transient current measurement was repeated on resonance with the \Phs~peak at the magnetic field of $B_0 = 350.3~\mathrm{mT}$ under application of microwave pulses with different length $\tau$.
\begin{figure}
\centering
\includegraphics[width=0.35\textwidth]{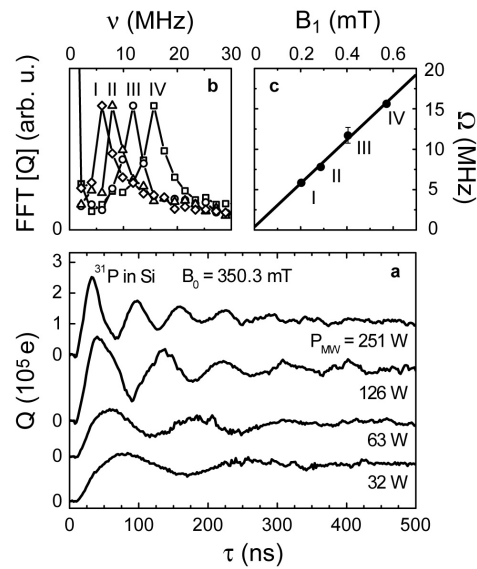}
\caption{Electrical observation of Rabi flops of \Phs~electron spins. (a) Photoconductivity change integrated from $t_1=7~\mu \mathrm{s}$ to $t_2=23~\mu \mathrm{s}$ after a coherent ESR
excitation as a function of the applied pulse length $\tau$ for
four different microwave powers $P_\mathrm{MW}$. (b) Plot of the Fast-Fourier Transform of $Q(\tau)$ for the four data sets shown in (a), normalized
to the intensity of the frequency peak. (c) Plot of the peak frequencies $\Omega$ from (b) as
a function of the microwave field amplitude $B_1$ exhibiting the linear relationship expected for Rabi flops.} \label{fig:wiggles}
\end{figure}
Figure~\ref{fig:wiggles}(a) displays the measurement of $Q(\tau)$
obtained for four different pulse powers $P_\mathrm{MW}$ under otherwise identical
experimental conditions as for the experiments shown in Fig.~\ref{fig:transients}.
We clearly see well developed oscillations which increase in frequency for increasing
microwave power. This can be seen more quantitatively in Fig.~\ref{fig:wiggles}(b), which displays the Fast-Fourier
Transform (FFT) of $Q(\tau)$. These curves were normalized to the maximum of their
respective frequency peaks. With increasing microwave field
strength $B_1 \propto \sqrt{P_{MW}}$, we observe a clear shift of the characteristic oscillation frequency $\Omega$ to
higher values. Fig.~\ref{fig:wiggles}(c) shows that the peak frequency
increases linearly with the excitation field $B_1$, as expected for Rabi oscillations. For the integration times used in Fig.~\ref{fig:wiggles}, a maximum of the recombining charge $Q(\tau)$ corresponds to a maximum of singlet states after the pulse. Since at low temperatures the spins of \Phs~and \Pbo~are in the $|\negmedspace \downarrow \downarrow \rangle$ triplet state before the pulse, the maximum of $Q$ indicates $|\negmedspace \uparrow \rangle$ of \Phs~after the pulse, a minimum corresponds to $|\negmedspace \downarrow \rangle$. Rabi echo experiments (not shown) reveal that the fast decay of the Rabi oscillations in Fig.~\ref{fig:wiggles}(a) is not due to incoherence but is caused by an inhomogeneous $B_1$ field which leads to a fast coherent dephasing of the spin ensemble.

On resonance, $\Omega = \gamma B_1$ should hold for isolated spin 1/2 states, where $\gamma=g\mu_B / \hbar$ is the gyromagnetic ratio and $\mu_B$ the Bohr magneton. From the linear fit of the
data displayed in Fig.~\ref{fig:wiggles}(c), we obtain a
proportionality factor of $\gamma_\mathrm{exp}=(1.03\pm 0.08)\gamma$. Since the ratio $\Omega / \gamma B_1$ is greater than 1 for strongly coupled systems~\cite{Rajevac:2006}, the experimentally observed $\Omega / \gamma B_1 \approx 1$ indicates a weakly coupled spin pair.

\begin{figure}
\centering
\includegraphics[width=0.35\textwidth]{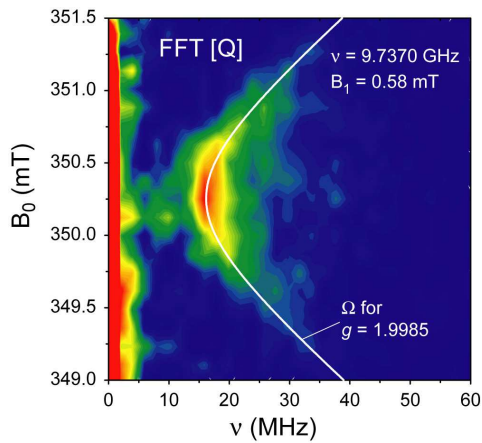}
\caption{Contour plot of the Fast-Fourier Transform of $Q(\tau)$ as a function of the applied
magnetic field $B_0$. The data confirm the predictions
for the off-resonance Rabi frequency $\Omega$ of a weakly-coupled spin-1/2 system with $g = 1.9985$ indicated by
the white solid line.} \label{fig:rabishift}
\end{figure}

Off resonance, the relation between the observed oscillation frequency $\Omega$ of
$Q(\tau)$ and the applied $B_1$ field is described by
the Rabi nutation frequency formula $\Omega = \sqrt{(\gamma B_1)^2
+ (\omega - \omega_L)^2}$ in the limit of uncoupled spins~\cite{Rabi:1937,Boehme:2003}. Here, $\omega$ is the angular
frequency of the microwave radiation and $\omega_L= \gamma B_0$ the Larmor
frequency of the spin. To test the Rabi frequency formula for spin-dependent recombination off resonance, $\Omega$ was measured  with a fixed $B_1$ and varied $B_0$ (or Larmor frequency $\omega_L$) near the high-field \Phs~resonance. Figure~\ref{fig:rabishift}
shows the results of these measurements in a contour plot of the FFT of $Q(\tau)$ as a funtion of $B_0$. A symmetric increase of $\Omega$ with increasing distance from the resonance peak is observed in perfect agreement with the predictions for $\Omega$ at $g = 1.9985$, indicated by the white line.
Therefore, the oscillations of $Q(\tau)$ shown in Figs.~\ref{fig:wiggles} and \ref{fig:rabishift} fully meet the predictions of Rabi oscillations of
weakly coupled spin 1/2 states.

The experiments presented here have been performed on ensembles of \Phs~donors. The charge noise achieved so far is about
$10^4~\mathrm{e}$ as shown in Fig.~\ref{fig:wiggles}. Since the flip of a single spin approximately leads to one elementary charge detected in the box-car pEDMR experiment~\cite{Boehme:2003}, the detection limit of pEDMR demonstrated here is around 11 orders of magnitude lower than for conventional nuclear magnetic resonance and about 7 orders of magnitude
lower than for ESR \cite{Bruker:1997}. The noise most likely is determined by the shot noise of shunt currents
due to excess charge carriers which diffuse into the c-Si bulk. Suppression of these currents via improved sample geometry e.g. using a buried oxide and lateral structuring for confinement should reduce the noise level, resulting in an even higher sensitivity. In fact, cw-EDMR was recently used to successfully detect as few as 100 \Phs~atoms implanted into intrinsic c-Si~\cite{McCamey:2006}.

The experimental results shown are an important step towards the realization of an electrical spin readout of donors. The experiments demonstrate (i) a spin-to-charge transfer mechanism required for \Phs~spin readout via \Phs-\Pbo~pairs and show that (ii) this spin pair is sufficiently weakly coupled so that the \Phs~hyperfine interaction can be observed, ultimately allowing the electrical detection of the nuclear spin state~\cite{Machida:2003}. The advantage of the concept is that it utilizes the native paramagnetic state at the c-Si/\sio~interface. The average distance of these \Pbo~centers can be adjusted between 10~nm and 100~nm by varying oxidation conditions. Given the localization length of the \Phs~donor~\cite{Koiller:2006}, it is therefore possible to prepare an interface where each \Phs~qubit interacts with one \Pbo~center.

This work was funded by Deutsche Forschungsgemeinschaft through SFB 631. The sample investigated was grown by G.~Vogg and F.~Bensch at Fraunhofer IZM in Munich.
\bibliography{electrical_P_rabi_flops}
\end{document}